\documentclass[english,showpacs,twocolumn,superscriptaddress,nofootinbib,prc]{revtex4-1}
\usepackage[T1]{fontenc}
\usepackage[latin9]{inputenc}
\usepackage{color}
\usepackage{babel}
\usepackage{float}
\usepackage{booktabs}
\usepackage{amsmath}
\usepackage{graphicx}
\usepackage[unicode=true,pdfusetitle,
 bookmarks=true,bookmarksnumbered=false,bookmarksopen=false,
 breaklinks=false,pdfborder={0 0 1},backref=false, colorlinks=true,linkcolor=red, citecolor=blue]
 {hyperref}

\makeatletter

\providecommand{\tabularnewline}{\\}

\makeatother

\begin{document}
\title{Averaged transverse momentum correlations of hadrons in relativistic
heavy-ion collisions}
\author{Yan-ting Feng}
\affiliation{School of Physics and Physical Engineering, Qufu Normal University,
Shandong 273165, China}
\author{Feng-lan Shao}
\email{shaofl@mail.sdu.edu.cn}

\affiliation{School of Physics and Physical Engineering, Qufu Normal University,
Shandong 273165, China}
\author{Jun Song }
\email{songjun2011@jnxy.edu.cn}

\affiliation{School of Physical Science and Intelligent Engineering, Jining University,
Shandong 273155, China}
\begin{abstract}
We compile experimental data for the averaged transverse momentum
($\left\langle p_{T}\right\rangle $) of proton, $\Lambda$, $\Xi^{-}$,
$\Omega^{-}$ and $\phi$ at mid-rapidity in Au+Au collisions at $\sqrt{s_{NN}}=$
200, 39, 27, 19.6, 11.5, 7.7 GeV and in Pb+Pb collisions at $\sqrt{s_{NN}}=$
2.76 TeV, and find that experimental data of these hadrons exhibit
systematic correlations. We apply a quark combination model with equal-velocity
combination approximation to derive analytic formulas of hadronic
$\left\langle p_{T}\right\rangle $ in the case of exponential form
of quark $p_{T}$ spectra at hadronization. We use them to successfully
explain the systematic correlations exhibited in $\left\langle p_{T}\right\rangle $
data of $p\Lambda$, $\Lambda\Xi^{-}$, $\Xi^{-}\Omega^{-}$ and $\Xi^{-}\phi$
pairs. We also use them to successfully explain the regularity observed
in $\left\langle p_{T}\right\rangle $ of these hadrons as the function
of $(dN_{ch}/dy)/(N_{part}/2)$ at mid-rapidity in central heavy-ion
collisions at both RHIC and LHC energies. Our results suggest that
the constituent quark degrees of freedom and the equal-velocity combination
of these constituent quarks at hadronization play important role in
understanding the production of baryons and $\phi$ meson at these
RHIC and LHC energies. 
\end{abstract}
\maketitle

\section{\emph{Introduction}\label{sec:Intro} }

In relativistic heavy-ion collisions, the hot nuclear matter is created
at early collision stage by intensively inelastic collisions of colliding
nucleons \citep{Heinz:2000bk,BRAHMS:2004adc,PHENIX:2004vcz,PHOBOS:2004zne,STAR:2005gfr,Gyulassy:2004zy}.
Subsequently, the matter expands, cools and finally decomposes into
hadrons scattering out. The evolution of hot nuclear matter is a complex
process governed by non-perturbative QCD and is mainly modeled by
hydrodynamic models \citep{Kolb:2003dz} and transport models \citep{Bass:1998ca,Bass:2000ib,Lin:2004en}
at present. Hadrons produced from hot nuclear matter always have certain
transverse momentum $p_{T}$, the component of momentum which is perpendicular
to the beam direction. The $p_{T}$ distribution of hadrons carries
lots of information on hot nuclear matter such as thermalization,
transverse collective flow generated by system expansion in both partonic
and hadronic stage, and is an important physical observable in experiments
of relativistic heavy-ion collisions. 

Rich experimental data for $p_{T}$ spectra of identified hadrons
at mid-rapidity are successively reported in heavy-ion collisions
at RHIC and LHC over the past decade \citep{PHENIX:2001vgc,PHENIX:2001hpc,STAR:2006uve,STAR:2008bgi,STAR:2008med,ALICE:2010mlf,ALICE:2013mez,ALICE:2014jbq,ALICE:2013cdo,ALICE:2013xmt,Adam:2019koz,STAR:2017sal}.
Based on these experimental data, lots of studies on properties of
hadronic $p_{T}$ distribution are carried out, which greatly improve
people's understandings on property of the created hot nuclear matter
and the mechanism of hadron production in relativistic heavy-ion collisions
\citep{Fries:2003vb,Greco:2003xt,Hwa:2002tu,Huovinen:2006jp,Chen:2006vc,Gyulassy:2003mc,Lokhtin:2005px,Zhang:2003wk,Tang:2008ud,Song:2011hk,Cassing:2008sv,Karpenko:2012yf}.
The average transverse momentum ($\left\langle p_{T}\right\rangle $)
of hadrons is obtained by integrating over $p_{T}$ spectra of hadrons.
It is dominated by property of hadronic $p_{T}$ spectra in the low
$p_{T}$ range and therefore it reflects the property of soft hadrons
and correspondingly that of hot nuclear matters. 

In this paper, we study the property of $\left\langle p_{T}\right\rangle $
of identified hadrons produced in relativistic heavy-ion collisions.
We compile experimental data for $\left\langle p_{T}\right\rangle $
of $\phi$, proton, $\Lambda$, $\Xi^{-}$ and $\Omega^{-}$ at mid-rapidity
in Au+Au collisions at $\sqrt{s_{NN}}=$ 200, 39, 27, 19.6, 11.5,
7.7 GeV and in Pb+Pb collisions at $\sqrt{s_{NN}}=$ 2.76 TeV. We
search the regularity in $\left\langle p_{T}\right\rangle $ data
of these hadrons and, in particular, their dependence on hadron species
and collision energy. We discuss what the underlying physics is responsible
to the observed regularity. In particular, we study the effect of
hadronization by an equal-velocity combination (EVC) mechanism of
quarks and antiquarks \citep{Song:2017gcz,Song:2020kak} in explaining
the experimental data of $\left\langle p_{T}\right\rangle $. We derive
analytic expression for $\left\langle p_{T}\right\rangle $ of identified
hadrons in EVC mechanism so as to give clear quark flavor dependence
of hadronic $\left\langle p_{T}\right\rangle $ and provide intuitive
explanations of experimental data. 

The paper is organized as follows. In Sec.~\ref{sec:EVC_model},
we briefly introduce our EVC model and derive $\left\langle p_{T}\right\rangle $
of identified hadrons for the simplified quark distributions at hadronization.
In Sec.~\ref{sec:pt_corr_hadron}, we show our finding for the systematic
correlation among experimental data for $\left\langle p_{T}\right\rangle $
of hadrons in relativistic heavy-ion collisions and give an intuitive
explanation using our EVC model. In Sec.~\ref{sec:pt_vs_nch}, we
show the regularity on $\left\langle p_{T}\right\rangle $ of hadrons
in central heavy-ion collisions as the function of $\left(dN_{ch}/dy\right)/\left(0.5N_{part}\right)$.
In Sec.~\ref{sec:decay}, we discuss the influence of resonance decays
on $\left\langle p_{T}\right\rangle $ correlations of hadrons. Finally,
the summary and discussion are given in Sec.~\ref{sec:Summary}. 

\section{\emph{$\left\langle p_{T}\right\rangle $ of hadrons in EVC model}
\label{sec:EVC_model}}

In this section, we apply a particular quark combination model \citep{Song:2017gcz,Gou:2017foe}
to describe the production of hadrons at hadronization and derive
$\left\langle p_{T}\right\rangle $ of different hadrons. The model
assumes the constituent quarks and antiquarks as the effective degrees
of freedom for the final parton system created in collisions at hadronization
stage. Based on constituent quark model of internal structure of hadrons
at low energy scale, the model assumes that the equal velocity combination
of these constituent quarks and antiquarks is main feature of hadron
formation at hadronization. The quark masses are taken as the constituent
masses so that the equal velocity combination of these constituent
quarks and antiquarks can correctly construct the on-shell hadron.
At mid-rapidity (i.e., taking $y=0$), $p_{T}$ distribution of a
hadron ($dN/dp_{T}$) is the product of those of (anti-)quarks 
\begin{align}
f_{B_{i}}\left(p_{T}\right) & =\kappa_{B_{i}}f_{q_{1}}\left(x_{1}p_{T}\right)f_{q_{2}}\left(x_{2}p_{T}\right)f_{q_{3}}\left(x_{3}p_{T}\right),\label{eq:fbi}\\
f_{M_{i}}\left(p_{T}\right) & =\kappa_{M_{i}}f_{q_{1}}\left(x_{1}p_{T}\right)f_{\bar{q}_{2}}\left(x_{2}p_{T}\right).\label{eq:fmi}
\end{align}
Here, moment fraction $x_{i}=m_{i}/(m_{1}+m_{2}+m_{3})$ ($i=1,2,3)$
in baryon formula satisfies $x_{1}+x_{2}+x_{3}=1$ and $x_{i}=m_{i}/(m_{1}+m_{2})$
($i=1,2)$ in meson formula satisfies $x_{1}+x_{2}=1$. $m_{i}$ is
constituent mass of quark $q_{i}$ and we take $m_{u}=m_{d}=0.3$
GeV and $m_{s}=0.5$ GeV. Coefficients $\kappa_{B_{i}}$ and $\kappa_{M_{i}}$
are independent of $p_{T}$, see Refs. \citep{Song:2020kak} for their
detailed expressions, and therefore are not involved in derivation
of averaged transverse momentum $\left\langle p_{T}\right\rangle $
in the following text.

The value of $\left\langle p_{T}\right\rangle $ is dominated by the
$p_{T}$ spectrum of particles in the low $p_{T}$ range. Therefore,
here we focus on $p_{T}$ spectra of quarks and antiquarks with low
$p_{T}$. Unfortunately, quarks of low $p_{T}$ is governed by non-perturbative
QCD dynamics and their distributions are difficult to calculate from
first principle. Considering that experimental data for $p_{T}$ spectra
of hadrons at mid-rapidity in the low $p_{T}$ range in relativistic
heavy-ion collisions are generally well fitted by exponential function
and/or Boltzmann distribution \citep{ALICE:2014jbq,ALICE:2013mez,STAR:2008bgi,STAR:2008med,STAR:2017sal,Adam:2019koz},
in this paper we take the following parameterization for quark $p_{T}$
spectra at mid-rapidity 
\begin{equation}
f_{q_{i}}\left(p_{T}\right)=\mathcal{N}p_{T}^{k}\exp\left[-\frac{\sqrt{p_{T}^{2}+m_{i}^{2}}}{T_{i}}\right],\label{eq:fq_thermal}
\end{equation}
which is convenient to derive analytic results of hadronic $\left\langle p_{T}\right\rangle $.
Here, $\mathcal{N}$ is coefficient to quantity the number of $q_{i}$,
which is irrelevant to $\left\langle p_{T}\right\rangle $ calculations.
$T_{i}$ is the slope parameter to quantify the exponential decrease
of the spectrum. Exponent $k$ tunes the behavior of the spectrum
at small $p_{T}$. In the case of two-dimensional Boltzmann distribution
in the rest frame we have $k=1$, and in the one-dimensional case
we have $k=0$. If we directly apply Eq.~(\ref{eq:fq_thermal}) to
fit the experimental data of $p_{T}$ spectra of hadrons by Eq.~\ref{eq:fbi},
we should take $k\approx1/3$ to properly describe baryon and $k\approx1/2$
to properly describe meson (mainly $\phi$). In addition, the effect
of strong collective radial flow should be included in the quark spectrum
in the laboratory frame, which is dependent on collision energies
in relativistic heavy-ion collisions. With these considerations, we
take $k$ as an relatively-free parameter in the range $[0,1]$ in
this study of hadronic $\left\langle p_{T}\right\rangle $ in relativistic
heavy-ion collisions at RHIC and LHC energies . 

\begin{figure*}[!htb]
\centering{}\includegraphics[width=0.95\textwidth]{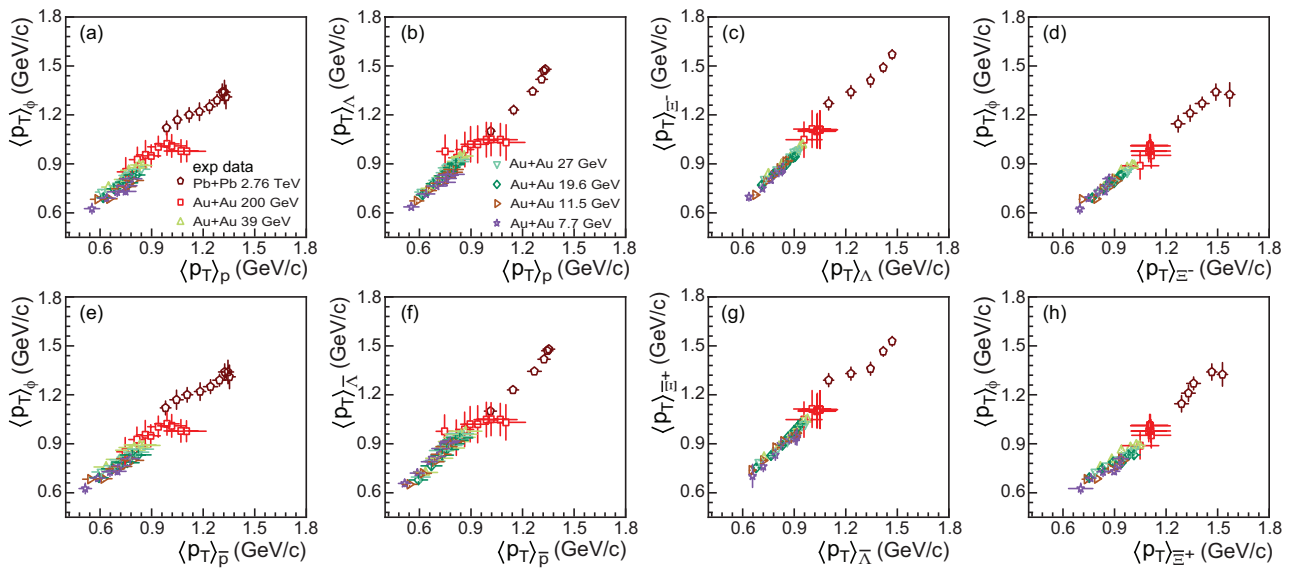}\caption{$\left\langle p_{T}\right\rangle $ of hadrons at mid-rapidity in relativistic heavy-ion collisions at different collision energies and collision centralities. Symbols are experimental data \citep{Adam:2019koz,ALICE:2013cdo,ALICE:2013xmt,ALICE:2014jbq,ALICE:2013mez,Estienne:2004di,STAR:2008bgi,STAR:2008med,STAR:2017sal} with quadratic combination of statistical and systematic uncertainties.  \label{fig:hapt_corr}}
\end{figure*}

Substituting Eq. (\ref{eq:fq_thermal}) into Eqs. (\ref{eq:fbi})
and (\ref{eq:fmi}), we obtain 
\begin{widetext}
\begin{equation}
\left\langle p_{T}\right\rangle _{B}=\frac{\int f_{B}\left(p_{T}\right)p_{T}dp_{T}}{\int f_{B}\left(p_{T}\right)dp_{T}}=\frac{\int p_{T}^{3k+1}\exp\left[-\left(\frac{x_{1}}{T_{1}}+\frac{x_{2}}{T_{2}}+\frac{x_{3}}{T_{3}}\right)\sqrt{p_{T}^{2}+m_{B}^{2}}\right]dp_{T}}{\int p_{T}^{3k}\exp\left[-\left(\frac{x_{1}}{T_{1}}+\frac{x_{2}}{T_{2}}+\frac{x_{3}}{T_{3}}\right)\sqrt{p_{T}^{2}+m_{B}^{2}}\right]dp_{T}},
\end{equation}
\begin{equation}
\left\langle p_{T}\right\rangle _{M}=\frac{\int f_{M}\left(p_{T}\right)p_{T}dp_{T}}{\int f_{M}\left(p_{T}\right)dp_{T}}=\frac{\int p_{T}^{2k+1}\exp\left[-\left(\frac{x_{1}}{T_{1}}+\frac{x_{2}}{T_{2}}\right)\sqrt{p_{T}^{2}+m_{M}^{2}}\right]dp_{T}}{\int p_{T}^{2k}\exp\left[-\left(\frac{x_{1}}{T_{1}}+\frac{x_{2}}{T_{2}}\right)\sqrt{p_{T}^{2}+m_{M}^{2}}\right]dp_{T}}
\end{equation}
where $m_{B}=m_{q_{1}}+m_{q_{2}}+m_{q_{3}}$ and $m_{M}=m_{q_{1}}+m_{\bar{q}_{2}}$.
We use the integral formula
\begin{equation}
\int_{0}^{\infty}p_{T}^{n}\exp\left[-a\sqrt{p_{T}^{2}+m^{2}}\right]dp_{T}=m^{n+1}\frac{2^{n/2}\Gamma(\frac{n+1}{2})}{\sqrt{\pi}}\frac{K_{n/2+1}(\alpha)}{\alpha^{n/2}}
\end{equation}
where $\alpha=a\,m$, $\Gamma(z)$ is Gamma function and $K_{n}(z)$
is the modified Bessel function of the second kind. We obtain
\begin{align}
\left\langle p_{T}\right\rangle _{B} & =\left(m_{q_{1}}+m_{q_{2}}+m_{q_{3}}\right)\sqrt{\frac{2}{\alpha_{B}}}\frac{\Gamma(3k/2+1)}{\Gamma(3k/2+\frac{1}{2})}\frac{K_{3k/2+3/2}(\alpha_{B})}{K_{3k/2+1}(\alpha_{B})},\label{eq:apt_B}
\end{align}
\begin{align}
\left\langle p_{T}\right\rangle _{M} & =\left(m_{q_{1}}+m_{\bar{q}_{2}}\right)\sqrt{\frac{2}{\alpha_{M}}}\frac{\Gamma(k+1)}{\Gamma(k+\frac{1}{2})}\frac{K_{k+3/2}(\alpha_{M})}{K_{k+1}(\alpha_{M})}\label{eq:apt_M}
\end{align}
\end{widetext}
with 
\begin{equation}
\alpha_{B}=\frac{m_{q_{1}}}{T_{1}}+\frac{m_{q_{2}}}{T_{2}}+\frac{m_{q_{3}}}{T_{3}}=\alpha_{q_{1}}+\alpha_{q_{2}}+\alpha_{q_{3}},\label{eq:alpha_B}
\end{equation}
 and 
\begin{equation}
\alpha_{M}=\frac{m_{q_{1}}}{T_{1}}+\frac{m_{\bar{q}_{2}}}{T_{2}}=\alpha_{q_{1}}+\alpha_{\bar{q}_{2}}.\label{eq:alpha_M}
\end{equation}
As shown by these expressions, $\left\langle p_{T}\right\rangle $
of different hadrons is correlated by the simple combination of slope
parameter $\alpha_{q_{i}}$ of quarks at hadronization.

Bessel function $K_{\nu}(\alpha)$ and Gamma function $\Gamma(z)$
usually have complex expressions. Here, we present the numerical approximations
for $\left\langle p_{T}\right\rangle _{B}$ and $\left\langle p_{T}\right\rangle _{M}$
\begin{align}
    &\left\langle p_{T}\right\rangle _{B}  \nonumber \\     
    &\approx\left(m_{q_{1}}+m_{q_{2}}+m_{q_{3}}\right)\left(0.26+0.024k+\frac{0.96+2.99k}{\alpha_{B}}\right),\label{eq:apt_B_approx}\\
    &\left\langle p_{T}\right\rangle _{M}  \approx\left(m_{q_{1}}+m_{\bar{q}_{2}}\right)\left(0.25+0.03k+\frac{0.97+1.99k}{\alpha_{M}}\right)\label{eq:apt_M_approx}
\end{align}
in order to see their dependence on $\alpha$ and $k$ in a numerically
intuitive way. The relative errors of two approximations are less
than about $3\%$ for the physical range of $\left\langle p_{T}\right\rangle _{B}$
and $\left\langle p_{T}\right\rangle _{M}$ in heavy-ion collisions
at RHIC and LHC energies studied in this paper. 

\section{correlations among $\left\langle p_{T}\right\rangle $ of different
hadrons\label{sec:pt_corr_hadron}}

In this section, we study the correlation among $\left\langle p_{T}\right\rangle $
of different hadrons. In Fig. \ref{fig:hapt_corr} (a), we present
$\left\langle p_{T}\right\rangle $ of proton as horizontal axis and
$\left\langle p_{T}\right\rangle $ of $\phi$ at correspondingly
collision energy and centrality as the vertical axis to study the
correlation between them. As we know, the mass of proton is close
to that of $\phi$ but the quark flavor composition of proton ($uud$)
is completely different from that of $\phi$ ($s\bar{s}$). Except
a few datum points in central Au+Au collisions at $\sqrt{s_{NN}}=200$
GeV, we see an relatively stable correlation between $\left\langle p_{T}\right\rangle $
of proton and that of $\phi$. In Fig. \ref{fig:hapt_corr} (b) and
(c), we show correlations among proton, $\Lambda$ and $\Xi^{-}$
in the manner of successive strangeness. In Fig. \ref{fig:hapt_corr}
(d), we show the correlation between $\phi$ and $\Xi^{-}$ which
both have two strange (anti-)quarks. We see the systematic correlations
among these hadrons. In Fig. \ref{fig:hapt_corr} (e-h), we show the
correlation among $\left\langle p_{T}\right\rangle $ data of $\phi$
and anti-baryons and also find systematic correlations among them.

As we know, the hot nuclear matters created in heavy-ion collisions
at different collision energies and centralities have different size
and geometry, different evolution times in partonic phase and in the
subsequent hadronic re-scattering stage, etc. The correlations shown
in Fig.~\ref{fig:hapt_corr} seem to assign these difference into
a systematic manner and therefore may indicate some underlying physics
which is universal in heavy-ion collision at both RHIC and LHC energies.
We think that the universal hadronization mechanism may be a possible
physical reason.

\begin{figure*}[!bht]
\centering{}\includegraphics[width=0.6\linewidth]{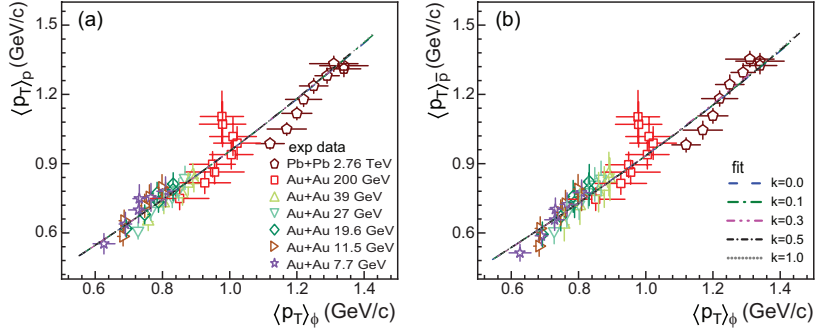}
    \caption{The fitted correlations between $\left\langle p_{T}\right\rangle $ of proton and $\phi$ (a) and that between $\left\langle p_{T}\right\rangle $ of anti-proton and $\phi$ (b). Symbols are experimental data \citep{ALICE:2014jbq,ALICE:2013mez,STAR:2008bgi,STAR:2008med,Adam:2019koz,STAR:2017sal}.  Lines with different types are fitting results by Eqs.~(\ref{eq:alpha_u_s}) and (\ref{eq:alpha_ubar_s}) with the corresponding coefficient values in Table. \ref{tab1}. \label{fig:alpha_us_p_phi}}
\end{figure*}

Therefore, we apply the EVC model in Sec.~\ref{sec:EVC_model} to
understand the above correlations in experimental data of hadronic
$\left\langle p_{T}\right\rangle $. Hadrons in Fig.~\ref{fig:hapt_corr}
are all made up of up, down, strange quarks and their antiquarks.
In our model, exponent parameter $k$ and slope parameters $\alpha_{u}$,
$\alpha_{d}$, $\alpha_{s}$, $\alpha_{\bar{u}}$, $\alpha_{\bar{d}}$
and $\alpha_{\bar{s}}$ are needed to fix in order to calculate $\left\langle p_{T}\right\rangle $
of hadrons according to Eqs.~(\ref{eq:apt_B}-\ref{eq:alpha_M}).
Here, we can assume the isospin symmetry between up and down quarks
$\alpha_{u}=\alpha_{d}$ at mid-rapidity in relativistic heavy-ion
collisions. We also assume the charge conjugation symmetry for strange
quarks and antiquarks $\alpha_{s}=\alpha_{\bar{s}}$, which is found
to be a good approximation at mid-rapidity in relativistic heavy-ion
collisions at RHIC and LHC energies \citep{Song:2020kak}. Finally,
only three slope parameters $\alpha_{u}$, $\alpha_{\bar{u}}$ and
$\alpha_{s}$ are left in addition to the exponent parameter $k$. 

If we know the correlation between $\alpha_{u}$ and $\alpha_{s}$,
we can calculate correlations among $\left\langle p_{T}\right\rangle $
of baryons and $\phi$, and therefore use them to explain experimental
data in Fig. \ref{fig:hapt_corr}(a-d). Applying Eqs.~(\ref{eq:apt_B})
and (\ref{eq:apt_M}) to proton and $\phi$, we have 
\begin{align}
\left\langle p_{T}\right\rangle _{p} & =3m_{u}\sqrt{\frac{2}{\alpha_{p}}}\frac{\Gamma(3k/2+1)}{\Gamma(3k/2+\frac{1}{2})}\frac{K_{3k/2+3/2}(\alpha_{p})}{K_{3k/2+1}(\alpha_{p})},\label{eq:apt_p}\\
\left\langle p_{T}\right\rangle _{\phi} & =2m_{s}\sqrt{\frac{2}{\alpha_{\phi}}}\frac{\Gamma(k+1)}{\Gamma(k+\frac{1}{2})}\frac{K_{k+3/2}(\alpha_{\phi})}{K_{k+1}(\alpha_{\phi})},\label{eq:apt_phi}
\end{align}
where
\begin{align}
\alpha_{p} & =2\alpha_{u}+\alpha_{d}=3\alpha_{u},\label{eq:alpha_proton}\\
\alpha_{\phi} & =\alpha_{s}+\alpha_{\bar{s}}=2\alpha_{s}\label{eq:alpha_phi}
\end{align}
according to Eqs. (\ref{eq:alpha_B}) and (\ref{eq:alpha_M}). By
fitting experimental data of $\left\langle p_{T}\right\rangle $ of
proton and $\phi$ shown in Fig.~\ref{fig:hapt_corr}(a), we can
reversely extract the correlation between $\alpha_{u}$ and $\alpha_{s}$,
which can be parameterized as 
\begin{equation}
\alpha_{u}(\alpha_{s})=c_{0}+c_{1}\alpha_{s}\label{eq:alpha_u_s}
\end{equation}
where $c_{0}$ and $c_{1}$ are two coefficients. Because the extraction
is also dependent on exponent parameter $k$, we list values of $c_{0}$
and $c_{1}$ at several different $k$ in Table.~\ref{tab1}. The
resulting fitted correlations between proton and $\phi$ with these
different extractions are shown as lines with different types in Fig.~\ref{fig:alpha_us_p_phi}
(a). In order to reduce to the bias in choice of $k$, we let these
different extractions to represent the same correlation between $\left\langle p_{T}\right\rangle $
of proton and $\phi$, i..e, these lines are coincident with each
other. By fitting experimental data of $\left\langle p_{T}\right\rangle $
of anti-proton and $\phi$, we also obtain the correlation between
$\alpha_{\bar{u}}$ and $\alpha_{s}$ with the parameterized form
\begin{equation}
\alpha_{\bar{u}}(\alpha_{s})=d_{0}+d_{1}\alpha_{s}.\label{eq:alpha_ubar_s}
\end{equation}
The values of coefficients $d_{0}$ and $d_{1}$ at several $k$ are
shown in Table. \ref{tab1} and the corresponding fit is shown in
Fig.~\ref{fig:alpha_us_p_phi} (b). At low RHIC energies where $\alpha_{s}$
is large, $\alpha_{\bar{u}}$ is different from $\alpha_{u}$ to a
certain extent, which is because the finite baryon density at low
collision energies will cause the asymmetry between up/down quarks
and their antiquarks. 

\begin{table}[H]

\centering{}\caption{Coefficients in Eqs.~(\ref{eq:alpha_u_s}) and (\ref{eq:alpha_ubar_s})
at different $k$, which are extracted from experimental data of $\left\langle p_{T}\right\rangle $
of (anti-)proton and $\phi$ at mid-rapidity in heavy-ion collisions
at RHIC and LHC energies. \label{tab1}}
\begin{tabular*}{6.5cm}{@{\extracolsep{\fill}}ccccc}
\toprule 
$k$ & $c_{0}$ & $c_{1}$ & $d_{0}$ & $d_{1}$\tabularnewline
\midrule
\midrule 
0.0 & -0.05 & 0.69 & -0.05 & 0.71\tabularnewline
\midrule 
0.1 & -0.07 & 0.76 & -0.07 & 0.78\tabularnewline
\midrule 
0.3 & -0.11 & 0.84 & -0.11 & 0.86\tabularnewline
\midrule 
0.5 & -0.15 & 0.89 & -0.15 & 0.91\tabularnewline
\midrule 
1.0 & -0.23 & 0.94 & -0.25 & 0.97\tabularnewline
\bottomrule
\end{tabular*}
\end{table}

\begin{figure*}[!htb]
\centering{}\includegraphics[width=0.95\linewidth]{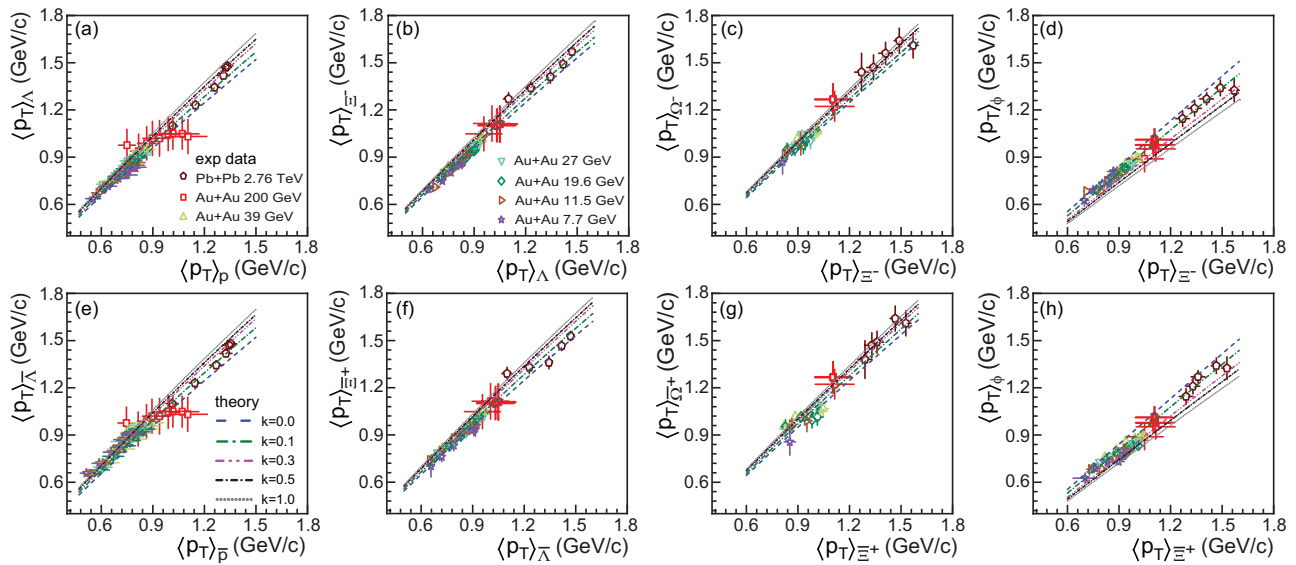}\caption{Correlations among $\left\langle p_{T}\right\rangle $ of hadrons at mid-rapidity in relativistic heavy-ion collisions at different collision energies and collision centralities. Symbols are experimental data \citep{ALICE:2013cdo,ALICE:2013xmt,ALICE:2014jbq,ALICE:2013mez,Estienne:2004di,STAR:2008bgi,STAR:2008med,STAR:2017sal,Adam:2019koz} and lines with different types are theoretical results with parameter values in Table. \ref{tab1}. \label{fig:hapt_corr_evc}}
\end{figure*}

With these two relationship, we can calculate correlations among $\left\langle p_{T}\right\rangle $
of various hadrons by Eqs.~(\ref{eq:apt_B}) and (\ref{eq:apt_M})
with Eqs.~(\ref{eq:alpha_proton}), (\ref{eq:alpha_phi}) and 
\begin{align}
\alpha_{\Lambda} & =2\alpha_{u}+\alpha_{s},\label{eq:alpha_Lambda}\\
\alpha_{\Xi} & =2\alpha_{s}+\alpha_{u},\label{eq:alpha_Xi}\\
\alpha_{\Omega} & =3\alpha_{s}\label{eq:alpha_Omega}
\end{align}
and the corresponding anti-baryons according to Eqs.~(\ref{eq:alpha_B})
and (\ref{eq:alpha_M}). In Fig. \ref{fig:hapt_corr_evc}, we present
theoretical results for $\left\langle p_{T}\right\rangle $ correlations
of $p\Lambda$, $\Lambda\Xi^{-}$, $\Xi^{-}\Omega$ and $\Xi^{-}\phi$
pairs, and these of anti-baryons, and we compare them with experimental
data. Theoretical results at different $k$ with corresponding coefficients
in Table. \ref{tab1} are shown as lines with different types. These
lines are different to a certain extent, which shows the theoretical
uncertainties due to the selection of exponent parameter $k$. Overall,
we see that the systematic feature of the correlations exhibited by
$\left\langle p_{T}\right\rangle $ data of these hadrons can be described
by our theoretical model. 

This result is quite interesting. Here we only consider the effect
of hadronization by EVC mechanism without any considerations on other
dynamical ingredients such as system size and geometry, evolution
time, hadronic re-scattering, etc. We run the event generators URQMD
3.4 \cite{Bass:1998ca} and AMPT 2.26 (1.26) \cite{Lin:2004en} which practically include those dynamical processes
and we do not find better description on systematic correlations in
Fig.~\ref{fig:hapt_corr} when we use default parameter values of
event generators. Therefore, our results in Fig.~\ref{fig:hapt_corr_evc}
indicate the important role of hadronization by EVC mechanism in describing
$\left\langle p_{T}\right\rangle $ of those hadrons in relativistic
heavy-ion collisions at both RHIC and LHC energies. 

\section{$\left\langle p_{T}\right\rangle $ of hadrons as the function of
$\left(dN_{ch}/dy\right)/\left(0.5N_{part}\right)$\label{sec:pt_vs_nch}}

Experimental data for hadronic $\left\langle p_{T}\right\rangle $
shown in the form of Fig.~\ref{fig:hapt_corr} reveal correlations
among $\left\langle p_{T}\right\rangle $ of different hadrons, which
are mainly relevant to hadronization mechanism according to our studies
in the previous section. In this section, we study another aspect
of $\left\langle p_{T}\right\rangle $ of hadrons, i.e., their absolute
values, and search some regularity underlying these experimental data
in relativistic heavy-ion collisions at RHIC and LHC energies. 

There are many physical ingredients that influence $\left\langle p_{T}\right\rangle $
of hadrons. Generally speaking, there are two main sources of generating
the transverse momentum of hadrons. The first source is the intensive
parton interactions at early collision stage which form the thermal
bulk nuclear matter and generate primordial thermal or stochastic
momentum of particles. Another source is the expansion of hot nuclear
matter in both partonic phase (if exists) and hadronic phase, which
generates the collective radial flow and therefore strengths the $\left\langle p_{T}\right\rangle $
of hadrons. The effects of two sources are both influenced by collision
parameters such as collision energy, collision centrality and collision
system. These collision parameters influence the size and geometry
of the bulk nuclear matter, the intensity of soft parton/particle
interactions, the time of system expansion and correspondingly the
magnitude of collective radial flow. In view of these complex ingredients,
it seems to be difficult to find a simple and perfect regularity for
$\left\langle p_{T}\right\rangle $ of hadrons by directly analyzing
the experimental data of $\left\langle p_{T}\right\rangle $ of hadrons
in relativistic heavy-ion collisions. 

\begin{figure*}[!htb]
\centering{}\includegraphics[width=0.95\linewidth]{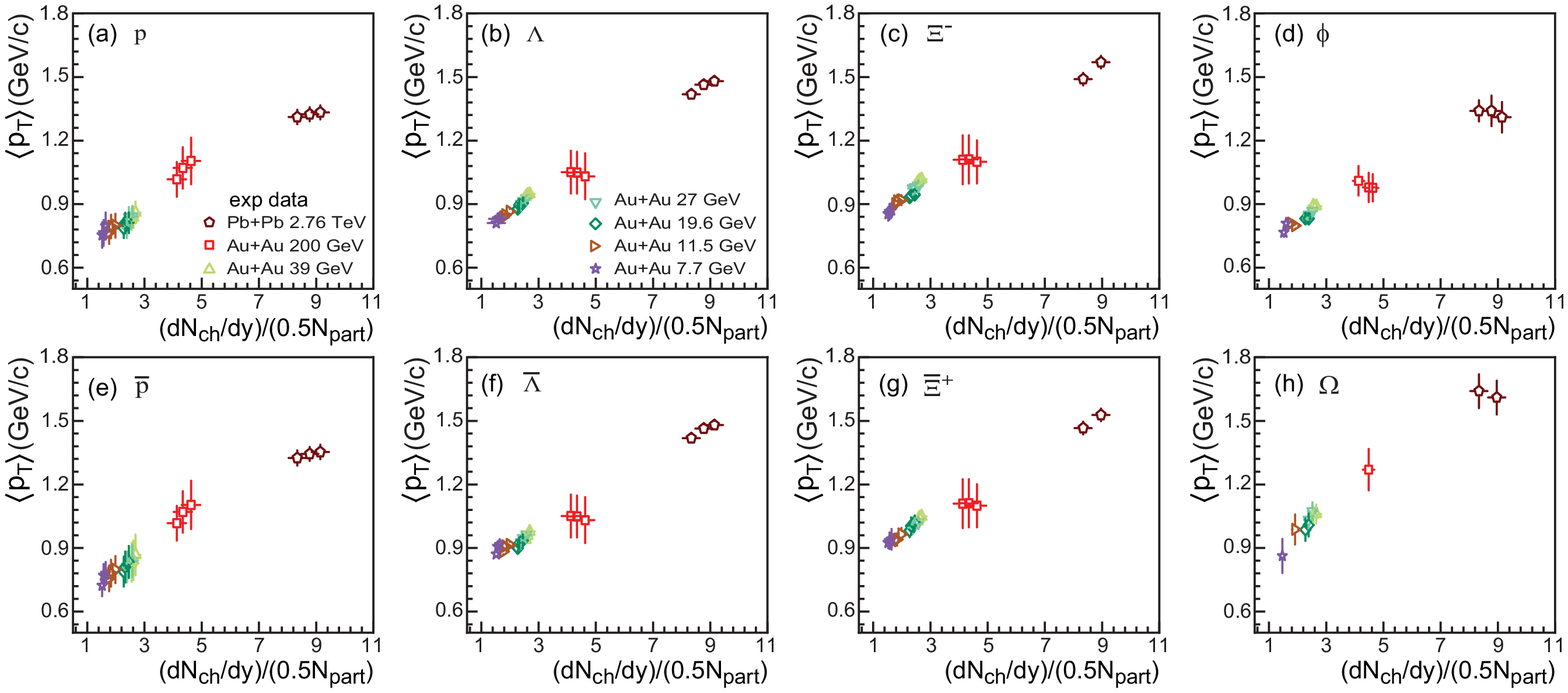}\caption{$\left\langle p_{T}\right\rangle $ of hadrons as the function of $\left(dN_{ch}/dy\right)/\left(0.5N_{part}\right)$ at mid-rapidity in central  heavy-ion collisions at different collision energies.  Symbols are experimental data \citep{ALICE:2013cdo,ALICE:2013xmt,ALICE:2014jbq,ALICE:2013mez,Estienne:2004di,STAR:2008bgi,STAR:2008med,STAR:2017sal,Adam:2019koz,ALICE:2010mlf}.
\label{fig:hapt_vs_dNchdy}}
\end{figure*}

Here, we try to take $\left(dN_{ch}/dy\right)/\left(0.5N_{part}\right)$
to quantify the excitation of hadronic $\left\langle p_{T}\right\rangle $.
$dN_{ch}/dy$ is the rapidity density of charged particles at mid-rapidity.
It can characterize the size of the created hot nuclear matter. In
general case, i.e., as other conditions (such as collision energy
and collision system) are not changed, the larger system means more
intensive particle excitation (i.e.,~higher stochastic momentum or
higher temperature) and more expansion (i.e.,~more radial flow).
Experimental observation have shown that $\left\langle p_{T}\right\rangle $
of hadrons generally positively responds to the $dN_{ch}/dy$ at given
collision energy and collision system \citep{Olimov:2021bjf,Petrovici:2017izo,ALICE:2014jbq,ALICE:2013mez,Estienne:2004di,STAR:2008bgi,STAR:2008med,STAR:2017sal,Adam:2019koz}.
Therefore, we take $dN_{ch}/dy$ as the main relevant ingredients
parameterizing $\left\langle p_{T}\right\rangle $ of hadrons. $N_{part}$
is the number of participant nucleons calculated in Glauber model
\citep{Miller:2007ri}, which depends on collision energy and collision
system and impact parameter. It can characterize the total amount
of energy deposited in the collision region and therefore characterize
the initial size and energy density of the created nuclear matter.
The ratio $\left(dN_{ch}/dy\right)/\left(0.5N_{part}\right)$ quantifies
the average number of charged particles produced by a pair of participant
nucleons. It can roughly characterize the average number of charged
particles produced by an ``unit'' effective energy deposited by
the collision of a pair of nucleons. In general, the higher $\left(dN_{ch}/dy\right)/\left(0.5N_{part}\right)$
means more intensive particle excitation which needs more intensive
parton interactions and also means more momentum generation. Therefore,
we expect $\left(dN_{ch}/dy\right)/\left(0.5N_{part}\right)$ should
positively correlate with $\left\langle p_{T}\right\rangle $ of hadrons. 

The geometry property of hot nuclear matter, mainly controlled by
impact parameter, also influence $\left\langle p_{T}\right\rangle $
of hadrons. In particular, in peripheral collisions where impact parameter
is large, various-order anisotropic flows are generated and will influence
the $\left\langle p_{T}\right\rangle $ to a certain extent by, for
example, asymmetric distribution of $p_{x}$ and $p_{y}$. Therefore,
in order to remove the freedom of impact parameter which is quite
complex to parameterize, we only use experimental data of hadronic
$\left\langle p_{T}\right\rangle $ in central collisions to search
their possible regularity with respect to $\left(dN_{ch}/dy\right)/\left(0.5N_{part}\right)$.

In Fig.~\ref{fig:hapt_vs_dNchdy}, we compile experimental data for
$\left\langle p_{T}\right\rangle $ of $\phi$ and (anti-)baryons
at mid-rapidity in central heavy-ion collisions at different collision
energies. We see that these data of hadronic $\left\langle p_{T}\right\rangle $
exhibit a clear regularity when we plot them as the function $\left(dN_{ch}/dy\right)/\left(0.5N_{part}\right)$.
Here data of $dN_{ch}/dy$ at mid-rapidity and $N_{part}$ are taken
from Refs. \citep{ALICE:2013mez,ALICE:2010mlf,STAR:2008med,STAR:2017sal}.
In calculation of $\left(dN_{ch}/dy\right)/\left(0.5N_{part}\right)$,
only experimental uncertainties of $dN_{ch}/dy$ are included. 

\begin{figure*}[!htb]
\centering{}\includegraphics[width=0.95\linewidth]{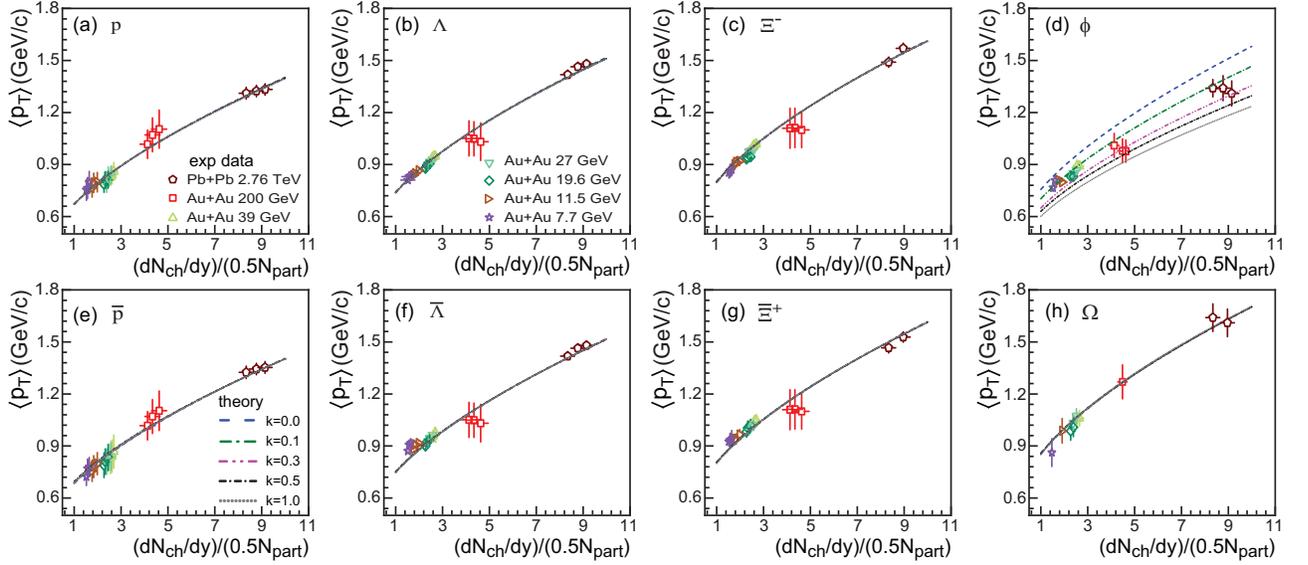}\caption{$\left\langle p_{T}\right\rangle $ of hadrons the function of $\left(dN_{ch}/dy\right)/\left(0.5N_{part}\right)$
at mid-rapidity in central heavy-ion collisions at different collision
energies. Symbols are experimental data \citep{ALICE:2013cdo,ALICE:2013xmt,ALICE:2014jbq,ALICE:2013mez,Estienne:2004di,STAR:2008bgi,STAR:2008med,STAR:2017sal,Adam:2019koz,ALICE:2010mlf}
and lines with different types are theoretical results with parameter
values in Table. \ref{tab2}. \label{fig:hapt_vs_dNchdy-qcm}}
\end{figure*}

According to behavior of experimental data in Fig.~\ref{fig:hapt_vs_dNchdy}
and the approximated formula of hadronic $\left\langle p_{T}\right\rangle $
in Eqs. (\ref{eq:apt_B_approx}) and (\ref{eq:apt_M_approx}), we
parameterize the $\left(dN_{ch}/dy\right)/\left(0.5N_{part}\right)$
dependence of slope parameter $\alpha_{q}$ of quarks as the following
form
\begin{align}
\alpha_{q} & =\left[g_{q}+h_{q}\left(\frac{dN_{ch}/dy}{N_{part}/2}\right)^{2/3}\right]^{-1}.\label{eq:alpha_q_nch}
\end{align}
Coefficients $g$ and $h$ of $u$, $\bar{u}$ and $s$ quarks can
be fixed by using Eqs.(\ref{eq:apt_B}) and (\ref{eq:apt_M}) to fit
experimental data of proton, anti-proton and $\Omega^{-}+\bar{\Omega}^{+}$.
Values of $g$ and $h$ at different $k$ are shown in Table.~\ref{tab2}.
These fittings to experimental data of proton, anti-proton and $\Omega^{-}+\bar{\Omega}^{+}$
are shown in Fig.~\ref{fig:hapt_vs_dNchdy-qcm} (a), (e) and (h)
as lines of different types. In order to avoid the bias in selection
of exponent parameter $k$, we let these different fitting groups
generate the same $\left(dN_{ch}/dy\right)/\left(0.5N_{part}\right)$
dependence for $\left\langle p_{T}\right\rangle $ of proton, anti-proton
and $\Omega^{-}+\bar{\Omega}^{+}$, that is, lines of different types
are coincident with each other in Fig.~\ref{fig:hapt_vs_dNchdy-qcm}
(a), (e) and (h). This treatment can enable us to study theoretical
uncertainty in prediction of other hadrons. Results for $\left\langle p_{T}\right\rangle $
of other hadrons as the function of $\left(dN_{ch}/dy\right)/\left(0.5N_{part}\right)$
are shown in Fig.~\ref{fig:hapt_vs_dNchdy-qcm}(b-d) and (f-g) and
are compared with experimental data. 

\begin{table}[H]
\centering{}\caption{Coefficients in Eq.~(\ref{eq:alpha_q_nch}) at different $k$, which
are extract from experimental data of $\left\langle p_{T}\right\rangle $
of (anti-)proton and $\Omega^{-}$ at mid-rapidity in heavy-ion collisions
at RHIC and LHC energies. \label{tab2}}
\begin{tabular*}{6.5cm}{@{\extracolsep{\fill}}ccccccc}
\toprule 
 & \multicolumn{2}{c}{$u$} & \multicolumn{2}{c}{$\bar{u}$} & \multicolumn{2}{c}{$s$}\tabularnewline
\midrule 
$k$ & $g$ & $h$ & $g$ & $h$ & $g$ & $h$\tabularnewline
\midrule
\midrule 
0.0 & 0.88 & 0.68 & 0.93 & 0.67 & 0.60 & 0.46\tabularnewline
\midrule 
0.1 & 0.64 & 0.52 & 0.69 & 0.52 & 0.42 & 0.36\tabularnewline
\midrule 
0.3 & 0.42 & 0.36 & 0.45 & 0.35 & 0.26 & 0.25\tabularnewline
\midrule 
0.5 & 0.30 & 0.27 & 0.34 & 0.27 & 0.19 & 0.19\tabularnewline
\midrule 
1.0 & 0.18 & 0.17 & 0.20 & 0.17 & 0.11 & 0.12\tabularnewline
\bottomrule
\end{tabular*}
\end{table}

We see that theoretical results of hyperons at different $k$ in Fig.~\ref{fig:hapt_vs_dNchdy-qcm}(b-c)
and (f-g) are coincident with each other and they are in good agreement
with experimental data. Results of $\phi$ in Fig.~\ref{fig:hapt_vs_dNchdy-qcm}(d)
are dependent on $k$ to a certain extent. In these results at different
$k$, we see that the results of $\phi$ at $k=0.1,0.3$ are globally
better than others. This feature is similar with that in correlations
of hadronic $\left\langle p_{T}\right\rangle $ in Fig.~\ref{fig:hapt_corr_evc}.

\section{Influence of resonance decay\label{sec:decay}}

In previous calculations, results of hadronic $\left\langle p_{T}\right\rangle $
are those for initially produced hadrons by hadronization and effects
of resonance decay are not yet included. For proton, $\Lambda$ and
$\Xi^{-}$, a certain fraction of these hadrons observed in experiments
comes from decay of higher-mass resonances such as those from octet
baryons $\Delta$, $\Sigma^{*}$ and $\Xi^{*}$, respectively. $\Omega^{-}$
and $\phi$ are generally expected to be less influenced. In this
section, we study the influence of resonance decay on the correlations
among $\left\langle p_{T}\right\rangle $ of different hadrons. 

We apply the quark combination model developed in previous works \citep{Song:2017gcz,Gou:2017foe}
to calculate the influence of resonance decay on $\left\langle p_{T}\right\rangle $
of hadrons. Following experimental corrections, results of $\Lambda$
and $\Xi^{-}$ do not include weak decay contributions but results
of proton and anti-proton include them. We adopt the following strategy
to quantify the effect of resonance decays. First, we use the model
to calculate the $\left\langle p_{T}\right\rangle $ of final-state
(anti-)proton and that of $\phi$ with the parameterized quark $p_{T}$
spectra in Eq.~(\ref{eq:fq_thermal}). We apply the model to fit
the $\left\langle p_{T}\right\rangle $ correlation between experimental
data of (anti-)protons and those of $\phi$ to obtain the correlation
between $\alpha_{u}$ and $\alpha_{s}$ with the parameterization
form Eq.~(\ref{eq:alpha_u_s}) and that between $\alpha_{\bar{u}}$
and $\alpha_{s}$ with Eq.~(\ref{eq:alpha_ubar_s}). The newly obtained
coefficients $c_{0}$, $c_{1}$, $d_{0}$, $d_{1}$ are slightly different
from those in Table \ref{tab1} due to the effect of resonance decays.
In the fitting process, we keep the same $p\phi$ and $\bar{p}\phi$
correlations shown as lines in Fig~\ref{fig:alpha_us_p_phi}. Second,
we calculate $\left\langle p_{T}\right\rangle $ correlations among
other hadron pairs and compare them with results in Fig. $\ref{fig:hapt_corr_evc}$
to study the effect of resonance decays. 

\begin{figure*}[!htb]
\centering{}\includegraphics[width=0.95\linewidth]{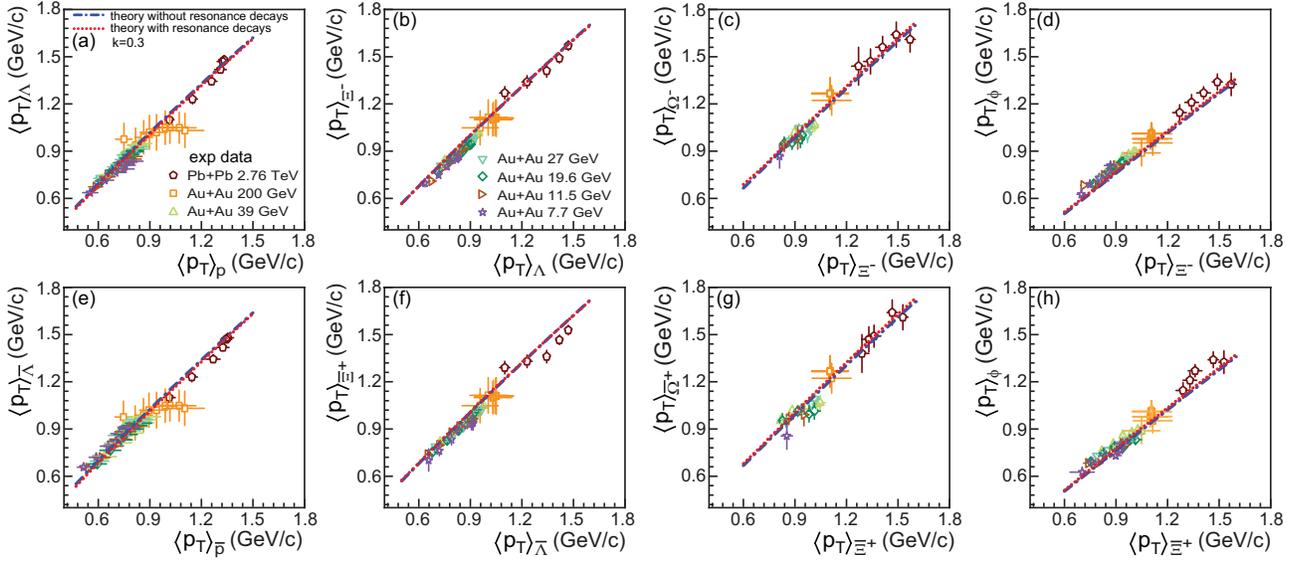}
    \caption{Correlations among $\left\langle p_{T}\right\rangle $ of hadrons at mid-rapidity in relativistic heavy-ion collisions at different collision energies and collision centralities. Symbols are experimental data \citep{ALICE:2013cdo,ALICE:2013xmt,ALICE:2014jbq,ALICE:2013mez,Estienne:2004di,STAR:2008bgi,STAR:2008med,STAR:2017sal,Adam:2019koz}.  The dashed lines are theoretical results including resonance decays and the dot-dashed lines are these not including resonance decays. \label{fig:hapt_corr_evc_final}}
\end{figure*}

In Fig.~\ref{fig:hapt_corr_evc_final}, we show $\left\langle p_{T}\right\rangle $
correlations among different final-state hadrons at given slope parameter
$k=0.3$ and compare them, dashed lines, with results of directly-produced
hadrons at same $k$, the dot-dashed lines, which are borrowed from
Fig.~\ref{fig:hapt_corr_evc}. Experimental data \citep{ALICE:2013cdo,ALICE:2013xmt,ALICE:2014jbq,ALICE:2013mez,Estienne:2004di,STAR:2008bgi,STAR:2008med,STAR:2017sal,Adam:2019koz}
are also presented. We see that two sets of results are only slightly
different, which indicates the weak influence of resonance decays
on $\left\langle p_{T}\right\rangle $ correlations of these hadrons.
Results at other values of $k$ are quite similar and therefore not
presented. 

\begin{figure*}[!htb]
\centering{}\includegraphics[width=0.95\linewidth]{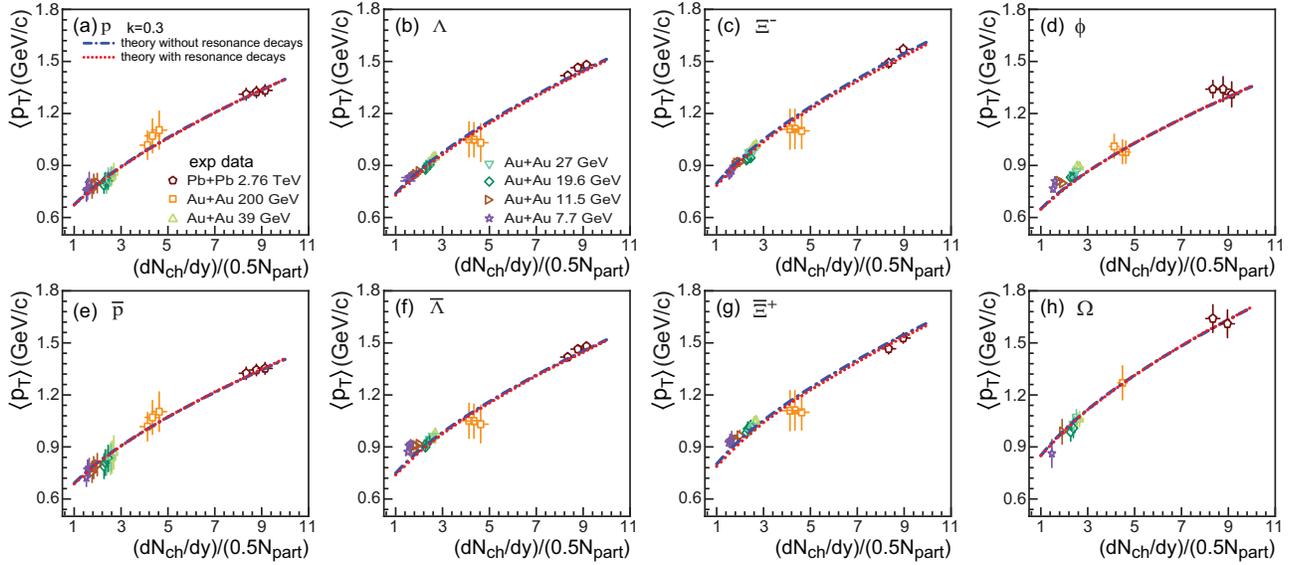}
    \caption{$\left\langle p_{T}\right\rangle $ of hadrons the function of $\left(dN_{ch}/dy\right)/\left(0.5N_{part}\right)$ at mid-rapidity in central heavy-ion collisions at different collision energies. Symbols are experimental data \citep{ALICE:2013cdo,ALICE:2013xmt,ALICE:2014jbq,ALICE:2013mez,Estienne:2004di,STAR:2008bgi,STAR:2008med,STAR:2017sal,Adam:2019koz,ALICE:2010mlf}.  The dashed lines are theoretical results including resonance decays and the dot-dashed lines are these not including resonance decays.  \label{fig:hapt_npart_evc_final}}
\end{figure*}

In the similar way, we further study the effect of resonance decays
on hadronic $\left\langle p_{T}\right\rangle $ as the function of
$\left(dN_{ch}/dy\right)/\left(0.5N_{part}\right)$ in central heavy-ion
collisions. In Fig.~\ref{fig:hapt_npart_evc_final}, the dashed lines
denotes results for $\left\langle p_{T}\right\rangle $ of hadrons
with resonance decays and dot-dashed lines denote results without
resonance decays. Symbols are experimental data \citep{ALICE:2013cdo,ALICE:2013xmt,ALICE:2014jbq,ALICE:2013mez,Estienne:2004di,STAR:2008bgi,STAR:2008med,STAR:2017sal,Adam:2019koz,ALICE:2010mlf}.
Experimental data of protons (a), anti-protons (e) and $\Omega^{-}$
(h) are used to determine parameters of (anti-)quarks at hadronization,
and we keep the same extent in reproducing these experimental data,
i.e., dashed lines are coincident with the dot-dashed lines in three
panels. In Fig.~\ref{fig:hapt_npart_evc_final} (b-d) and (f-g),
we see the difference between dashed lines and dot-dashed lines is
small, which indicate the weak influence of resonance decays on $\left\langle p_{T}\right\rangle $
correlations of these hadrons.

\section{Summary and discussion\label{sec:Summary}}

In this paper, we have applied a quark combination model with equal-velocity
combination approximation to study the averaged transverse momentum
($\left\langle p_{T}\right\rangle $) correlations of proton, $\Lambda$,
$\Xi^{-}$, $\Omega^{-}$ and $\phi$ in relativistic heavy-ion collisions.
We derived analytic formulas of hadronic $\left\langle p_{T}\right\rangle $
in the case of exponential form of quark $p_{T}$ spectra at hadronization,
which can clarify correlations among $\left\langle p_{T}\right\rangle $
of identified hadrons based on the constituent quark structure of
hadrons. We used these analytic formulas to explain the systematic
correlations exhibited in $\left\langle p_{T}\right\rangle $ data
of $p\Lambda$, $\Lambda\Xi^{-}$, $\Xi^{-}\Omega^{-}$ and $\Xi^{-}\phi$
pairs and those of anti-baryons. We discussed the regularity for $\left\langle p_{T}\right\rangle $
of these hadrons as the function of $(dN_{ch}/dy)/(N_{part}/2)$ at
mid-rapidity in central heavy-ion collisions at both RHIC and LHC
energies, and used our model to self-consistently explain $\left\langle p_{T}\right\rangle $
of these hadrons as the function of $(dN_{ch}/dy)/(N_{part}/2)$.
In these studies, we use experimental data of (anti-)protons to fix
the property of up/down (anti-)quarks and those of $\Omega$ or $\phi$
to fix that of strange quarks at hadronization. Then we predict correlations
among $\left\langle p_{T}\right\rangle $ of other hadron pairs and
compare with experimental data to test the theoretical consistency.
Moreover, we studied the effects of resonance decays on $\left\langle p_{T}\right\rangle $
correlations of hadrons and find they are weak in comparison with
hadronization. 

Our studies shown that the $\left\langle p_{T}\right\rangle $ correlations
among experimental data of proton, $\Lambda$, $\Xi^{-}$, $\Omega^{-}$
and $\phi$ in Au+Au collisions at RHIC energies and Pb+Pb collisions
at $\sqrt{s_{NN}}=2.76$ TeV can be self-consistently described by
the equal-velocity combination mechanism of constituent quarks and
antiquarks at hadronization. This indicates the important role of
constituent quarks and antiquarks as the effective degrees of freedom
of the hot nuclear matter at hadronization stage and their equal-velocity
combination as the main feature of hadron formation in relativistic
heavy-ion collisions. The current study is consistent with our previous
works in studying the elliptic flow of these hadrons and the quark
number scaling property of $p_{T}$ spectra of $\Omega^{-}$ and $\phi$
in relativistic heavy-ion collisions at RHIC and LHC energies using
the same quark combination mechanism \citep{Song:2019sez,Song:2021ygg}. 

In relativistic heavy-ion collisions, re-scatterings of hadrons after
hadronization will influence momentum of hadrons to a certain extent.
For example, the signal of $\phi$ may be lost by the scattering of
their decay daughters with the surrounding hadrons and $\phi$ may
be also generated by the coalescence of two kaon. We will study this
hadronic rescattering effect in the future works. In addition, we
will also carry out a systematic study on $p_{T}$ spectra of identified
hadrons at mid-rapidity in different centralities in Au+Au collisions
at STAR BES energies. $p_{T}$ spectra of identified hadrons contain
more dynamical information than their $\left\langle p_{T}\right\rangle $,
which can be used to further test our quark combination model at low
RHIC energies. 

\section{Acknowledgments}

We thank Prof. X. L. Zhu for providing us the experimental data of
hadronic $\left\langle p_{T}\right\rangle $ in Au+Au collisions at
RHIC energies. This work is supported in part by Shandong Provincial
Natural Science Foundation (ZR2019YQ06, ZR2019MA053), the National
Natural Science Foundation of China under Grant No. 11975011, and
Higher Educational Youth Innovation Science and Technology Program
of Shandong Province (2019KJJ010).

\bibliographystyle{apsrev4-1}
\bibliography{ref}

\end{document}